\documentclass[12pt,nofootinbib]{revtex4-1}
\usepackage{amsmath,amssymb, pxfonts}
\usepackage{graphicx}
\usepackage{ytableau}

\newcommand{\beq}{\begin{equation}}
\newcommand{\beqs}{\begin{equation*}}
\newcommand{\eeq}{\end{equation}}
\newcommand{\eeqs}{\end{equation*}}

\allowdisplaybreaks

\begin{document}


\title{Reconstructing spacetime from the  hologram, even in the classical limit, requires physics beyond the Planck scale.\\
\begin{small}
ÒEssay written for the Gravity Research Foundation 2016 Awards for Essays on Gravitation.Ó 
\end{small}}

\author{David Berenstein }\email{Corresponding author: dberens@physics.ucsb.edu} \author{Alexandra Miller }\email{apmiller@physics.ucsb.edu}\affiliation{ Department of Physics, University of California, Santa Barbara, CA 93106}

\begin{abstract}
We argue in this essay that for classical configurations of gravity in the AdS/CFT setup, it is in general impossible to reconstruct the bulk geometry from the leading asymptotic behavior of the classical 
fields in gravity alone.  This is possible sufficiently near the vacuum, but not more generally. We argue this by using a counter-example  that utilizes the supersymmetric geometries constructed by Lin, Lunin, and Maldacena. In the dual quantum field theory, the additional data required to complete the geometry is encoded in modes that near  the vacuum geometry lie beyond the Planck scale.
\end{abstract}

\maketitle


Understanding how quantum mechanics and gravity can be made compatible is one of the thorniest problems in theoretical physics. This is especially true in
light of the black hole information paradox \cite{Hawking:1974sw}. 
One of the main claims of the AdS/CFT correspondence \cite{Maldacena:1997re} is that it provides a definition of quantum gravity for spacetimes that are asymptotically of the form $AdS_{d+1}\times X$ in terms of a quantum field theory in $d$-dimensions that resides on the boundary of the AdS geometry.  In this setup, the information paradox is resolved in principle: the quantum field theory of the boundary does not violate quantum mechanics. We do not yet understand how the paradox is resolved in terms of the geometric variables of gravity.

The boundary theory is often referred to as the hologram, while the full higher dimensional geometry is called the bulk.
A rather important open problem in the study of this duality is the problem of reconstruction: how to derive the geometric data of the  bulk spacetime from the hologram. 
The hope is that if we understand this procedure sufficiently well, we might finally understand what was the {\em wrong assumption} in the original calculation by Hawking.

Here, we will discuss how much information is needed from the hologram in order to reconstruct the bulk spacetime, specifically when the spacetime is a classical solution of gravity. 
This problem is well understood near the vacuum state. Fields in the AdS spacetime are in one to one correspondence with certain families of local operators on the boundary \cite{Witten:1998qj}. The expectation values of these operators are related to the behavior of the solutions of the classical fields as they approach the boundary. Very small classical excitations in the bulk imprint themselves on these  expectation values of the hologram in a way that makes it  possible to reconstruct the bulk solution from the expectation values  in a perturbative expansion \cite{Hamilton:2005ju,Kabat:2011rz}. 

Does this work beyond a perturbative argument? In this essay, we will argue that the answer to this question is generically no. The way we will argue for this outcome is with a very concrete counter-example, where we can see the failure explicitly.
We will then argue that there is a mechanism in the quantum field theory that provides the additional data necessary to reconstruct the spacetime, but that this data is hidden 
in modes that  lie beyond the Planck scale from the point of view of the  vacuum geometry.

The counter-example can be constructed in the maximally supersymmetric theory with excited states that preserve as much supersymmetry as possible. The complete set of these solutions has been classified by Lin, Lunin and, Maldacena \cite{Lin:2004nb} and we will refer to them as LLM geometries. 
What makes this example special is that we also understand the complete set of such states in the dual field theory \cite{Corley:2001zk,Berenstein:2004kk}, so we can understand the problem  in detail not only in the geometric space of classical solutions, but also in the Hilbert space of the quantum theory.

The LLM geometries are completely specified by a single function $z(w, \bar w, y)$, which must obey a known  differential equation. They are regular and horizon free. Geometric regularity forces  $z(w, \bar w,0)$ to take one of two possible values, $\pm 1/2$.  This can be represented by a two-coloring of the complex $w, \bar w$ plane.  A single round disk will give rise to the $AdS_5\times S^5$ spacetime. These patterns have fixed area and are identical to the configuration space of a fixed amount of incompressible liquid in two dimensions. 

The coloring can be described by a step function $\rho(w, \bar w)$ that takes the value one in the region of one color and is zero otherwise. 
Given $\rho(w, \bar w)$, the solution for $z$ is 
\begin{equation}
z(w, \bar w, y) = \frac 12 - \frac {y^2}{\pi}\int \frac{\rho(w', \bar w') d^2 w'}{(|w -w'|^2+y^2)^2}\label{eq:soln}
\end{equation}

The boundary of the  LLM geometries  is located in the region where $r^2=(y^2+w\bar w)\to \infty$ and $z\to 1/2-N (r^2-w\bar w)/r^4+O(1/r^4)$. Indeed, $z$ admits an expansion
in powers of $w/r^2, \bar w/r^2, 1/r^2$. The expressions that arise are linear in the moments $M_{m,n} = \int \rho(w, \bar w) w^n \bar w^m d^2 w$ of $\rho (w, \bar w)$. As shown in the works \cite{Balasubramanian:2006jt,Skenderis:2007yb} (see also \cite{Balasubramanian:2007zt} for how to use some higher moments as conserved charges with which to distinguish quantum states), the problem of computing expectation values of fields on the boundary can be reduced to studying a particular set of these multipole moments.  Conversely, in this essay, we will ask if it is possible to determine  $\rho$ from only a subset of the moments.  
\begin{figure}
\begin{center}
\includegraphics[width=10 cm]{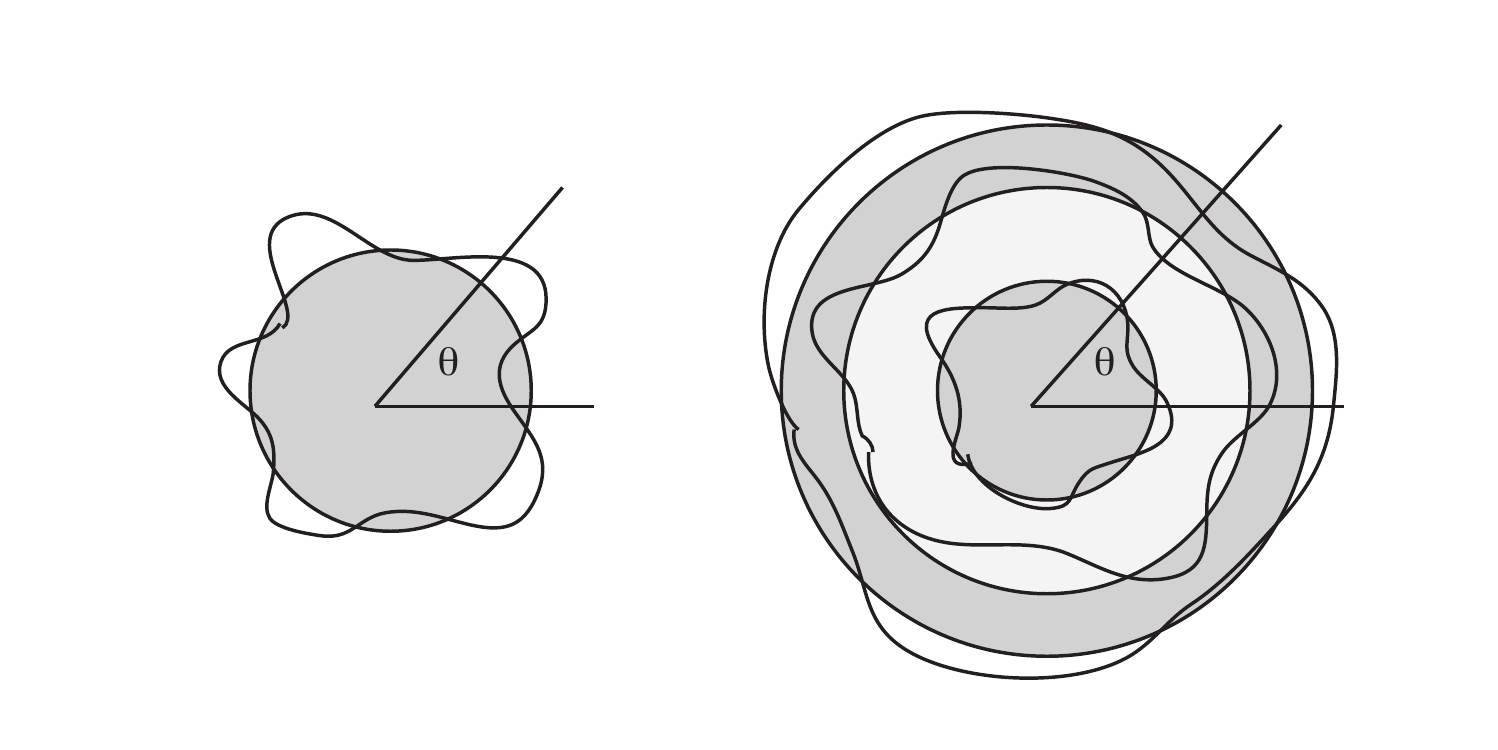}
\end{center}
\caption{An example of a single droplet, versus a multi-droplet geometry, deformed away from a setup of concentric circles. Notice that a ray starting from the origin intersects various deformations of the edges.}
\label{fig:multi-edge}
\end{figure}

To see that this is a reasonable possibility, consider a small deformation of the lowest energy state, which is represented by a circular disk of radius $r_0= \sqrt N$ (as shown in the left figure of \ref{fig:multi-edge}).  We want the edge to  be slightly deformed, so that in polar coordinates the edge of the droplet can be described by a radial function $r(\theta) = r_0 +\delta r(\theta)$. In these coordinates, $w= r\exp(i\theta)$ and because $\rho$ is just a step function, we can integrate along the radial direction to get
\begin{eqnarray}
 M_{0,m} &=&   \iint \rho(r,\theta) r^m \exp(-i m \theta)   r dr d\theta \\
 &=& N\delta_{m,0} + \int d\theta r_0^{m+1} \delta r(\theta) \exp(-i m\theta) d\theta+ O(\delta r^2) \label{eq:fourier}
\end{eqnarray} 
We see that to linear order, the moments $M_{m,0}$  become exactly the Fourier modes of $\delta r(\theta)$. 
Higher $M_{n,m}$ are also linear in these Fourier modes, so they are redundant.  We now see why it is reasonable to think that the subset of moments $M_{0,m}$ might be sufficient to calculate the function $z$ and hence the full geometry:  we need only to perform the inverse Fourier transform to build the geometry.  This procedure fails in general geometries.

Consider now the coloring illustrated on the right of figure \ref{fig:multi-edge}, which shows a droplet excitation with a deformed disk and annulus that are concentric. We can quantify the deformations by a set of functions, $\delta r_i(\theta)$ for each edge. One can then repeat the calculations that ended up with the Fourier coefficients \eqref{eq:fourier}.  However, in this case one would immediately see that the $M_{0,m}$ would only capture a particular linear combination of the three $\delta r_i$.  This problem increases as one considers geometries with more droplets.  We find then that it is impossible to reconstruct the geometry from the knowledge of this restricted data.

We can also compute the $M_{n,m}$ for the multi-droplet geometry and find that to linearized order they give a different linear combination of the Fourier modes of $\delta r_i$.  So, in general, to reconstruct the bulk we would need roughly one of the $M_{n,m}$ for each  mode of $\delta r_i$. Classically, the state can have an arbitrary number of such circles, so to solve the problem in the general case, we would need all of the $M_{n,m}$. This is the same as knowing the function $z$ (and therefore the full geometry) at the start. To reconstruct spacetime, we need to know it already.  We find that reconstruction in the general case is not only hard, but it is ambiguous: many geometries can have the same classical boundary data.  This is a counterexample to bulk reconstruction in the sense of \cite{Hamilton:2005ju}, because some fluctuations in the bulk are not coupled linearly to the expectation values that are available on the boundary.

Now, we will discuss the implications of this observation for the full quantum problem, rather than just the classical problem. The  difficulty in reconstructing the interior is not on first figuring out if the state preserves supersymmetry. The condition we need to satisfy  is that the charge of the state under one of  the rotations of the sphere  is equal to the energy of the state.  Both of these are readily measured on the boundary. The problem really lies in that we need more modes than are available on the vacuum to describe the excited state of the multi-edge droplet picture in figure \eqref{fig:multi-edge}, one mode for each of the Fourier modes of $\delta r_i(\theta)$. Where do these modes come from?

To better understand the physics of the boundary, we need to use the dictionary \cite{Witten:1998qj}. The single-particle states that preserve this amount of symmetry are massless in ten dimensions.   The list of modes can be read off from a table   \cite{Gunaydin:1984fk, Kim:1985ez}. Because these modes preserve a large amount of supersymmetry, they can be  followed from families of operators in the free field theory limit of the boundary theory as we increase the strength of interactions. This physics is  determined by the eigenvalues of a single matrix degree of  freedom in the quantum field theory  \cite{Berenstein:2004kk}. Let us say there are $N$ such eigenvalues. This means that we only need to know up to $N$ invariants of the matrix to compute them. This was argued to be a form of the stringy exclusion principle \cite{McGreevy:2000cw}.  These eigenvalues, when properly quantized,  actually determine quantum droplets of incompressible fluids in two dimensions.   The counting of excitations of the ground state droplet give a single tower of states that stops at mode $N$. This has fewer degrees of freedom than the classical gravitational theory, where the tower has no end,  but notice in the latter case we had assumed that $N\to\infty$ first.

Here is where the physics of the Planck scale enters.  A particle with momentum of order the Planck scale is not a mode with momentum $n=N$, but it is at much smaller values, around $n=N^{1/4}$. The vast majority of the modes that are required to describe the field theory data for such a supersymmetric state  lie beyond the Planck scale. 

To describe the degrees of freedom of the additional edges of the droplets, we need to borrow supersymmetric modes from the UV somehow, as these are the only other modes that preserve the correct amount of supersymmetry.  It is now becoming  clear why it was not possible to reconstruct the bulk with the classical theory: the physics responsible for the new modes of the geometry is way above the cutoff. In the strict classical limit, where $N$ approaches infinity,  this cutoff was sent to infinity first! 

A natural question is how something that started its life at short wavelengths (high energies) became effectively a mode that is present in the long wavelength limit of the theory in the excited state. The way this must happen is that the map in equation \eqref{eq:fourier} is non-linear. A non-linear combination of modes adds the frequencies and wavelengths
of the linear fields. To get a long wavelength (low frequency), there must be negative contributions as well as positive ones. The negative contributions in frequency indicate that the state is not in the ground state, but in an excited state. Bound states of large positive and large negative energy will have the required characteristics.

That  such bound states  play an important role implies that the modes are organized in a non-trivial way in the quantum wave function. This  requires the right type of   entanglement between the different UV modes. 
This entanglement  allows one to change the topology of spacetime creating a droplet configuration with multiple edges. These multi-edge geometries actually have different topologies of the spacetime \cite{Lin:2004nb}, thus realizing some of the ideas of Van Raamsdonk where topology changes of spacetime are related to entanglement \cite{VanRaamsdonk:2010pw}. 
It is natural to speculate that this is more generally true in gravity and that this paves a way to resolve the paradox of Hawking \cite{Hawking:1974sw}.  Transplanckian modes are doing something important.

\acknowledgements

We are very grateful to G. Horowitz and D. Marolf for many discussions. Work  supported in part by the department of Energy under grant {DE-SC} 0011702.


\begin{thebibliography}{99}

\bibitem{Hawking:1974sw} 
  S.~W.~Hawking,
 ``Particle Creation by Black Holes,''
  Commun.\ Math.\ Phys.\  {\bf 43}, 199 (1975)
  [Commun.\ Math.\ Phys.\  {\bf 46}, 206 (1976)].


\bibitem{Maldacena:1997re} 
  J.~M.~Maldacena,
  ``The Large N limit of superconformal field theories and supergravity,''
  Int.\ J.\ Theor.\ Phys.\  {\bf 38}, 1113 (1999)
  [Adv.\ Theor.\ Math.\ Phys.\  {\bf 2}, 231 (1998)]
  doi:10.1023/A:1026654312961
  [hep-th/9711200].
  
  
\bibitem{Witten:1998qj} 
  E.~Witten,
``Anti-de Sitter space and holography,''
  Adv.\ Theor.\ Math.\ Phys.\  {\bf 2}, 253 (1998)
  [hep-th/9802150].
  
\bibitem{Hamilton:2005ju} 
  A.~Hamilton, D.~N.~Kabat, G.~Lifschytz and D.~A.~Lowe,
``Local bulk operators in AdS/CFT: A Boundary view of horizons and locality,''
  Phys.\ Rev.\ D {\bf 73}, 086003 (2006)
  doi:10.1103/PhysRevD.73.086003
  [hep-th/0506118].
  
  
\bibitem{Kabat:2011rz} 
  D.~Kabat, G.~Lifschytz and D.~A.~Lowe,
  ``Constructing local bulk observables in interacting AdS/CFT,''
  Phys.\ Rev.\ D {\bf 83}, 106009 (2011)
  doi:10.1103/PhysRevD.83.106009
  [arXiv:1102.2910 [hep-th]].
  
  
\bibitem{Lin:2004nb} 
  H.~Lin, O.~Lunin and J.~M.~Maldacena,
``Bubbling AdS space and 1/2 BPS geometries,''
  JHEP {\bf 0410}, 025 (2004)
  doi:10.1088/1126-6708/2004/10/025
  [hep-th/0409174].
  
  
\bibitem{Corley:2001zk} 
  S.~Corley, A.~Jevicki and S.~Ramgoolam,
  ``Exact correlators of giant gravitons from dual N=4 SYM theory,''
  Adv.\ Theor.\ Math.\ Phys.\  {\bf 5}, 809 (2002)
  [hep-th/0111222].
  
\bibitem{Berenstein:2004kk} 
  D.~Berenstein,
  ``A Toy model for the AdS / CFT correspondence,''
  JHEP {\bf 0407}, 018 (2004)
  doi:10.1088/1126-6708/2004/07/018
  [hep-th/0403110].
  
  
\bibitem{Balasubramanian:2006jt} 
  V.~Balasubramanian, B.~Czech, K.~Larjo and J.~Simon,
  ``Integrability versus information loss: A Simple example,''
  JHEP {\bf 0611}, 001 (2006)
  doi:10.1088/1126-6708/2006/11/001
  [hep-th/0602263].
  
  
\bibitem{Skenderis:2007yb} 
  K.~Skenderis and M.~Taylor,
  ``Anatomy of bubbling solutions,''
  JHEP {\bf 0709}, 019 (2007)
  doi:10.1088/1126-6708/2007/09/019
  [arXiv:0706.0216 [hep-th]].
  
\bibitem{Balasubramanian:2007zt} 
  V.~Balasubramanian, B.~Czech, K.~Larjo, D.~Marolf and J.~Simon,
  ``Quantum geometry and gravitational entropy,''
  JHEP {\bf 0712}, 067 (2007)
  doi:10.1088/1126-6708/2007/12/067
  [arXiv:0705.4431 [hep-th]].
  
  
\bibitem{Gunaydin:1984fk} 
  M.~Gunaydin and N.~Marcus,
  ``The Spectrum of the s**5 Compactification of the Chiral N=2, D=10 Supergravity and the Unitary Supermultiplets of U(2, 2/4),''
  Class.\ Quant.\ Grav.\  {\bf 2}, L11 (1985).
  doi:10.1088/0264-9381/2/2/001
  
\bibitem{Kim:1985ez} 
  H.~J.~Kim, L.~J.~Romans and P.~van Nieuwenhuizen,
  ``The Mass Spectrum of Chiral N=2 D=10 Supergravity on S**5,''
  Phys.\ Rev.\ D {\bf 32}, 389 (1985).
  doi:10.1103/PhysRevD.32.389
  

\bibitem{McGreevy:2000cw} 
  J.~McGreevy, L.~Susskind and N.~Toumbas,
  ``Invasion of the giant gravitons from Anti-de Sitter space,''
  JHEP {\bf 0006}, 008 (2000)
  doi:10.1088/1126-6708/2000/06/008
  [hep-th/0003075].
  

  

  
  
\bibitem{VanRaamsdonk:2010pw} 
  M.~Van Raamsdonk,
  ``Building up spacetime with quantum entanglement,''
  Gen.\ Rel.\ Grav.\  {\bf 42}, 2323 (2010)
  [Int.\ J.\ Mod.\ Phys.\ D {\bf 19}, 2429 (2010)]
  doi:10.1007/s10714-010-1034-0, 10.1142/S0218271810018529
  [arXiv:1005.3035 [hep-th]].



  
\end{thebibliography}
\end{document}